# The Communication of Meaning in Anticipatory Systems: A Simulation Study of the Dynamics of Intentionality in Social Interactions

Loet Leydesdorff

*Amsterdam School of Communications Research (ASCoR), University of Amsterdam, Kloveniersburgwal 48, 1012 CX Amsterdam, The Netherlands; http://www.leydesdorff.net ; loet@leydesdorff.net*

**Abstract.** Psychological and social systems provide us with a natural domain for the study of anticipations because these systems are based on and operate in terms of intentionality. Psychological systems can be expected to contain a model of themselves and their environments; social systems can be strongly anticipatory and therefore co-construct their environments, for example, in techno-economic (co-)evolutions. Using Dubois' hyper-incursive and incursive formulations of the logistic equation, these two types of systems and their couplings can be simulated. In addition to their structural coupling, psychological and social systems are also coupled by providing meaning reflexively to each other's meaning-processing. Luhmann's distinctions among (1) interactions between intentions at the micro-level, (2) organization at the meso-level, and (3) self-organization of the fluxes of meaningful communication at the global level can be modeled and simulated using three hyper-incursive equations. The global level of self-organizing interactions among fluxes of communication is retained at the meso-level of organization. In a knowledge-based economy, these two levels of anticipatory structuration can be expected to propel each other at the supra-individual level.

## 1. INTRODUCTION

The hyper-incursive formulation of the logistic equation—$x_t = ax_{t+1}(1 - x_{t+1})$ [1]—provides us with an operationalization of the concept of "double contingency" which has been central to the theory of social systems [2-7]. According to Luhmann [5:70], "double contingency" can be considered as the auto-catalyst of social processes between reflexive individuals.

The concept of double contingency was first formulated by the Amercian sociologist Talcott Parsons, who defined "double contingency" in 1951 as follows:

> The expectation is not defined "*Being* what I am, alter's treatment of me must take one of the following alternatives" but "Depending on *which of several alternatives open to me I take*, I will set *alter* a problem to which he will react in terms of the alternative system of his own which is oriented to my action."[2:94]

According to Parsons [3: 436], the theory of games could be considered as a most sophisticated analysis of the implications of double contingency. However, one can also specify the dynamics of expectations in terms of the theory of strongly anticipatory systems [8,9] and provide the above (formal) equation with the following (substantive) interpretation: *Ego* (at $x_t$) operates on the basis of an expectation of its own next state ($x_{t+1}$) and the next state of an *Alter* $(1 - x_{t+1})$. Note that *Alter* is now defined in terms of *Ego*'s expectations; the relationship between expectations constructed in each human mind precedes a possible interaction between *Ego* and *Alter*.

While Parsons based his definition mainly on American pragmatism [2,9], the German sociologist Niklas Luhmann elaborated on double contingency in the continental tradition. He based himself on Edmund Husserl's transcendental phenomenology. Husserl had concluded that "intersubjectivity" provides us with an intentionality different from and transcendental to subjectivity [10:144]. While subjective intentionality is a natural consequence of the *cogito* of the *cogitantes*, intersubjective intentionality remains an uncertain *cogitatum*, that is, an inter-human construct of expectations. The two substances (or in Husserl's wordings "monads") are different to the extent that they can be considered as *analytically* independent [12,13].

Husserl noted that he had no instruments beyond a transcendental apperception of this new domain and therefore had to refrain from its empirical investigation [11:138]. Luhmann [6,14] made the communication of meaning among human beings the focus of his sociology. The communication of meaning relates to the theory of anticipatory systems since meaning is provided to information from the perspective of hindsight. At each moment of time, the communication and degree of codification of meaning can be measured, for example, by using semantic maps [15,16]. By providing meaning to the historical events from the perspective of hindsight, a reflexive system generates a model of these events or—more precisely—the expected information values contained in series of events. Such a model can be used by a reflexive mind for the weak anticipation [1,17,18].

When meanings can also be communicated between weakly anticipatory systems, a next-order layer of *cogitata* can be expected as a result. I shall argue that this construct can become strongly anticipatory, that is, co-constructing its own future states. For example, the techno-economic (co-)evolution in the social system reconstructs this system to the extent that more recently the (sub)dynamics of a knowledge-based economy has been constructed on top of the older—since informational—dynamics of markets [19,20].

While biological systems can already provide meaning to the information [17,21] the *differentia specifica* of social and psychological systems is this *processing* of meaning. Meaning-processing systems can codify meaning into symbolic meaning and develop discursive reasoning. I follow Luhmann's assumption [6,12,14] that meaning is communicated in social systems, but can be made conscious only by psychological systems.

Using the *incursive* formulation of the logistic equation [$x_t = ax_{t-1}(1 - x_t)$], the selection term $(1 - x_t)$ provides observers with a weakly anticipatory model [18]. The social system contains an additional anticipatory mechanism because of its distributedness [22,23]. When these two (analytically different) anticipatory mechanisms operate upon each other, a strongly anticipatory system can be among the results [20]. This system of meaning-processing self-organizes its knowledge base in terms of meanings which make a difference for the construction of its next stage.

## 2. DOUBLE CONTINGENCY AND HYPER-INCURSIVITY

Dubois' formula for hyper-incursivity [1] captures the second (anticipatory) dimension of the double contingency in general. However, each other agent (e.g., $y_t$) can be expected to maintain a similar set of expectations, but with potentially another parameter $a$ and consequentially another relation to the future of itself and its environment. How is the development of double contingency dependent on this "bifurcation parameter" $a$?

The equation can be rewritten as follows [1:9]:

$$x_t = ax_{t+1}(1 - x_{t+1}) \tag{1}$$

$$x_t = ax_{t+1} - ax_{t+1}^2$$

$$ax_{t+1}^2 - ax_{t+1} + x_t = 0$$

$$x_{t+1}^2 - x_{t+1} + x_t/a = 0$$

In general, the latter equation has two solutions [24]:

$$x_{t+1} = \tfrac{1}{2} \pm \tfrac{1}{2}\sqrt{[1 - (4/a)\, x_t]} \tag{2}$$

Given that $0 \leq x \leq 1$, the curve $x = 0.5 \pm 0.5\sqrt{(1 - (4/a))}$ for $x = 1$ sets limits to the possible values which can be reached by the hyper-incursive system. This line is penciled into Figure 1. In addition to the two domains of possible solutions of Equation 2 for $a \geq 4$, Figure 1 shows the bifurcation diagram of the logistic equation for $a < 4$, and the steady state of the incursive equation [$x_t = ax_{t-1}(1 - x_t)$; $x_t = x_{t-1} \rightarrow x = (a - 1)/a$] [24].

For $a \geq 4$, two sets of expectations are hyper-incursively generated at each time step depending on the plus or the minus sign in the equation. After $N$ time steps, $2^N$ future states would be possible. Thus, this social system of expectations continuously needs a mechanism for making decisions between options. Otherwise, the system would rapidly become overburdened with uncertainty [1:9]. The decisions are made in the layer of historically-rooted but weakly anticipatory systems.

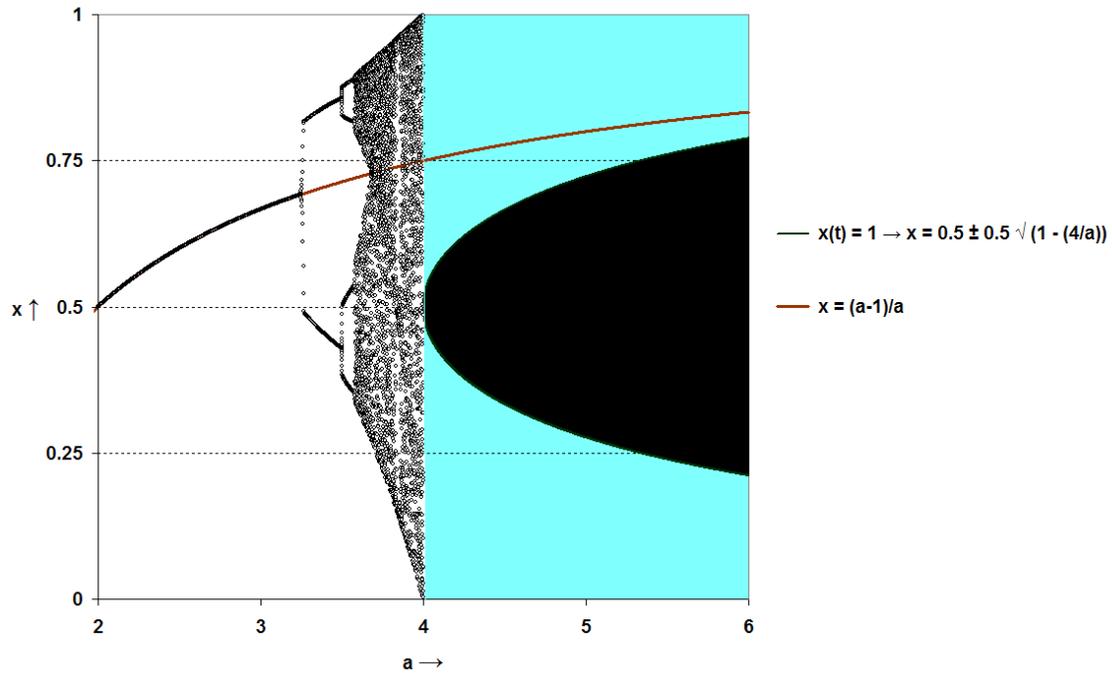

**FIGURE 1.** The social system as a result of hyper-incursion.

Note that the logistic equation and its hyper-incursive (but not the incursive) formulation are symmetrical in $t$: the logistic curve refers exclusively to $(t - 1)$ and the hyper-incursive equation to $(t + 1)$. Because of this symmetry in the formulas, both systems can be expected to reach a limit value for $a = 4$.

Can $a$ also have values lower than four in the strongly anticipatory case?

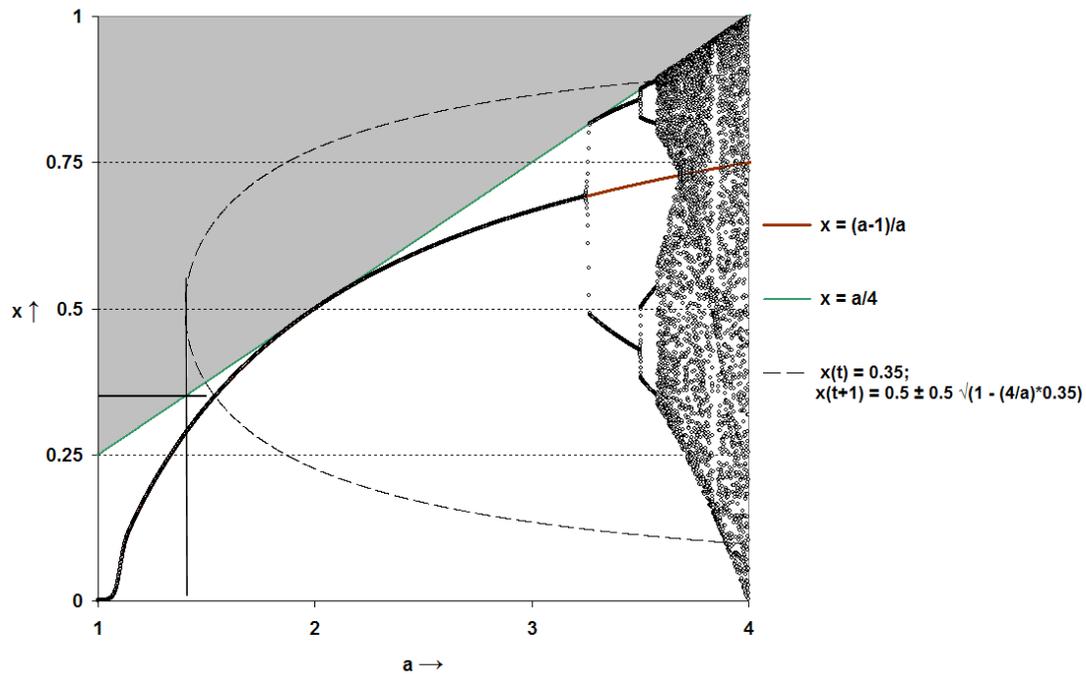

**FIGURE 2.** Possible penetrations of the social system into the biological variation ($a < 4$).

The term under the root in Equation 2 is positive for $x_t \leq a/4$: this condition is always met for $a \geq 4$, but sets a borderline to the possible penetrations of the hyper-incursive (social) system into the biological variation ($a < 4$). In Figure 2, this limitation is elaborated for $a = 1.2$ and $x_t = 0.35$. Since in a next step $x_{t+1}$ would become 0.5, and thus larger than $a/4$ (= 0.3), the strongly anticipatory system would in this case not be able to proceed with a next step to $x_{t+2}$. In general, the strongly anticipatory system can reconstruct the biological variation (e.g., technologically), but only for a relatively short period of time.[1]

## 3. INTENTIONALITY, IDENTITY, AND REFLECTION

Two more models can be derived from the logistic equation:

$$x_t = ax_t(1 - x_{t+1}) \tag{3}$$

$$x_t = ax_{t+1}(1 - x_t) \tag{4}$$

***Equation 3*** evolves into $x = (a - 1)/a$, that is, a constant for each value of $a$. This constant is equal to the steady state of the incursive equation [$x_t = ax_{t-1}(1 - x_t)$]. This steady state can be derived from the incursive formulation of the logistic equation as follows [24]:

$$x_{t+1} = ax_t(1 - x_{t+1}) \tag{3a}$$

$$x_{t+1} = ax_t - ax_t x_{t+1} \tag{3b}$$

$$x_{t+1}(1 + ax_t) = ax_t \tag{3c}$$

$$x_{t+1} = ax_t / (1 + ax_t) \tag{3d}$$

The steady state can be found by solving $x_t = x_{t+1}$ as follows:

$$x = ax / (1 + ax) \tag{5}$$

$$x(1 + ax) = ax \tag{5a}$$

$$ax^2 + (1 - a)x = 0 \tag{5b}$$

$$x = 0 \lor x = (a - 1)/a \tag{5c}$$

In other words, the incursive equation tends to converge on a steady state which can be considered as one of the available sub-dynamics of the strongly anticipatory system of meaning processing. The continuous line in Figure 1 represents the function $x = (a - 1)/a$.

This continuity allows the anticipatory system to maintain its identity for different values of $a$. Identity is based on the expectation of continuity of the "self" in the next stage of an anticipatory system while the aperture for the reflection—the parameter $a$—may vary. Note that institutions like individuals can be expected to develop an identity for organizing the communication of meaning in historical time.

***Equation 4*** evolves into $x_{t+1} = (1/a) [x_t / (1 - x_t)]$. This routine formalizes the reflexive operation: when $x_t > [a / (1 + a)]$ a pulse is generated which first overshoots the value of one (in a virtual domain of possible expectations), but then generates a negative value (Figure 3). Dubois (*personal communication*, July 16, 2008) noted that Eq. 4 can be derived as the *time inverse* of his [1] incursive (Pearl-Verhulst) equation $x_t = ax_{t-1}(1 - x_t)$.[2] However, the dynamics are different.

---

[1] For $a = 2$, the hyper-incursive system has a single ("natural"?) solution at $x_t = 0.5$. In the case of $a < 4$, there exists a basin of attraction of the for the fixed state $(a - 1)/a$ of Equation 2 with the plus sign, such that if $x_0$ is in this basin, and the hyperincursive decision rule allows the system to choose the plus equation, then $x_t$ will persist indefinitely. This never occurs when $a < 3$ (Marke Burke, *personal communication*, February 1, 2008.)

[2] The demonstration is as follows:

$$x_{t+1} = ax_t(1 - x_{t+1})$$

is equal to the following equation for dt = 1:

$$x_{t+dt} = ax_t(1 - x_{t+dt})$$

The time reverse of this equation is obtaind for dt → –dt, with the negative discrete time –dt:

$$x_{t-dt} = ax_t(1 - x_{t-dt})$$

or, with –dt = –1:

$$x_{t-1} = ax_t(1 - x_{t-1})$$

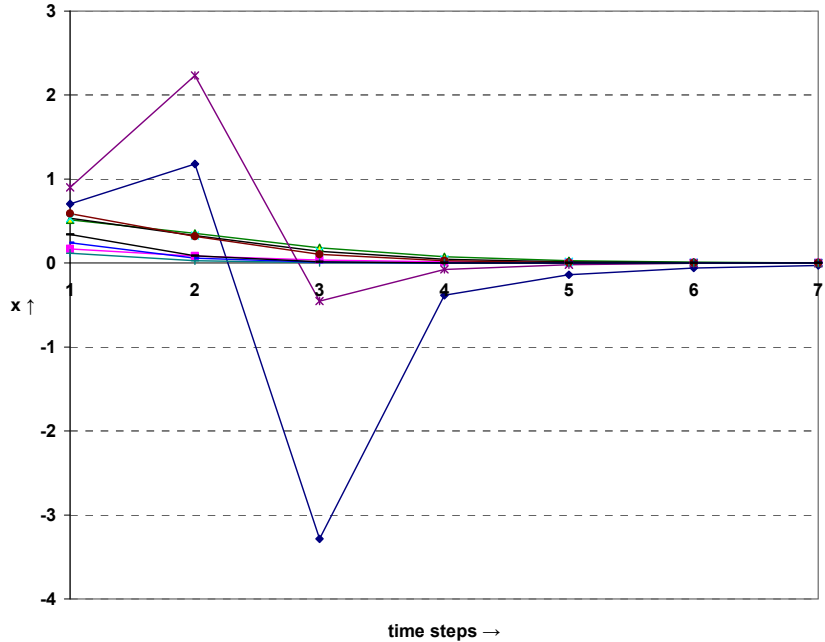

**FIGURE 3.** Simulation of Equation 4: the value of $x$ at $t = 1$ is drawn randomly ($a = 4$).[3]

The negative value provides a mirror image of a representation at a specific moment in time, and thus allows for the reflection. Reflection enables us to bounce a communication between the subdynamics of communication systems. At the level of the social system, this subdynamic formalizes the reflexive operation of providing meaning with reference to a horizon of possible meanings [25].

In summary, the three (hyper-)incursive equations which contain the future state in the argument, specify the systemic processes of meaning-processing: identity formation (Equation 3), reflexivity (Equation 4), and the mutual expectation of expectations in the doubly contingent relations between reflexive systems (Equation 1). While identity and reflexivity are more commonly defined in social theory [26], double contingency could hitherto not be operationalized. From the perspective of the theory of anticipatory systems, the second dimension of the double contingency in intentional systems—that is, the expectation of expectations—provides us with a natural system of strong anticipations.

## 4. DECISIONS AND HISTORICAL TRAJECTORIES

The hyper-incursive system (Equation 1) cannot further be developed (at $a \geq 4$) without decisions taken by reflexive agents (or institutional agency) because of the continuous production of uncertainty. A psychological system can be considered as the minimal unit of reflection for making choices [27,28]. If decisions are socially organized—for example, by using decision rules—an institutional layer can increasingly be shaped. This institutional layer of organization provides a retention mechanism and therefore a hold for a next round of interacting expectations at the supra-individual level [29,30].

In other words, the social system is dually layered as a forward-moving and information-processing retention mechanism versus sets of possible expectations which flow as intentions through the networks. The fluxes contain selections on Husserl's "horizons of meaning" [25:89f.]. These possible horizons of meaning are not given, but continuously undergoing reconstruction while weakly anticipatory systems make decisions [31]. In other words, these systems co-evolve in two layers of strong and weak anticipation, respectively. When the communication becomes more complex, the communicative competencies of the carriers of the communication are challenged to

---

So with a time translation of t → t +1 on the whole equation, one obtains:

$$x_t = ax_{t+1}(1 - x_t) \tag{4}$$

[3] For $a = 4$, the pulse is generated for values of $x_t > 0.8$.

improve their capacity by learning. The communication can further be developed when layers of codification can be sustained. Knowledge, for example, can be considered as a meaning that makes a difference; the exchange of knowledge as discursive knowledge requires more codification than the communication of meaning itself.

Because the various equations are derived from the same logistic equation, the steady state of the weakly anticipatory system is equal to an available state of the strongly anticipatory system. However, the strongly anticipatory system contains more subdynamics than the weakly anticipatory one. Unlike individuals, the social system is not continuously tending towards integration, but can also operate in a distributed mode and further develop its differentiation into subsystems. When the social system is more integrated—as in a high culture—the system can be stable over long periods of time; the incursive routines prevail and tend towards a steady state. When the social system is more differentiated, the hyper-incursive equation begins to play a role because the subsystems can entertain expectations about the expectations entertained in other subsystems and thus generate a double contingency, that is, a strongly anticipatory mechanism at the level of the social system.

In summary, the social system is complex and therefore composed of several subdynamics [21]. When, among the various subdynamics, the hyper-incursive routine prevails, pressure to take decisions increases. This can be experienced subjectively as time pressure; the system becomes less predictable because it increasingly co-constructs its own future states.[4] This social system can be considered as "non-trivial" [33] and unexpected consequences of decisions can then increasingly be expected.

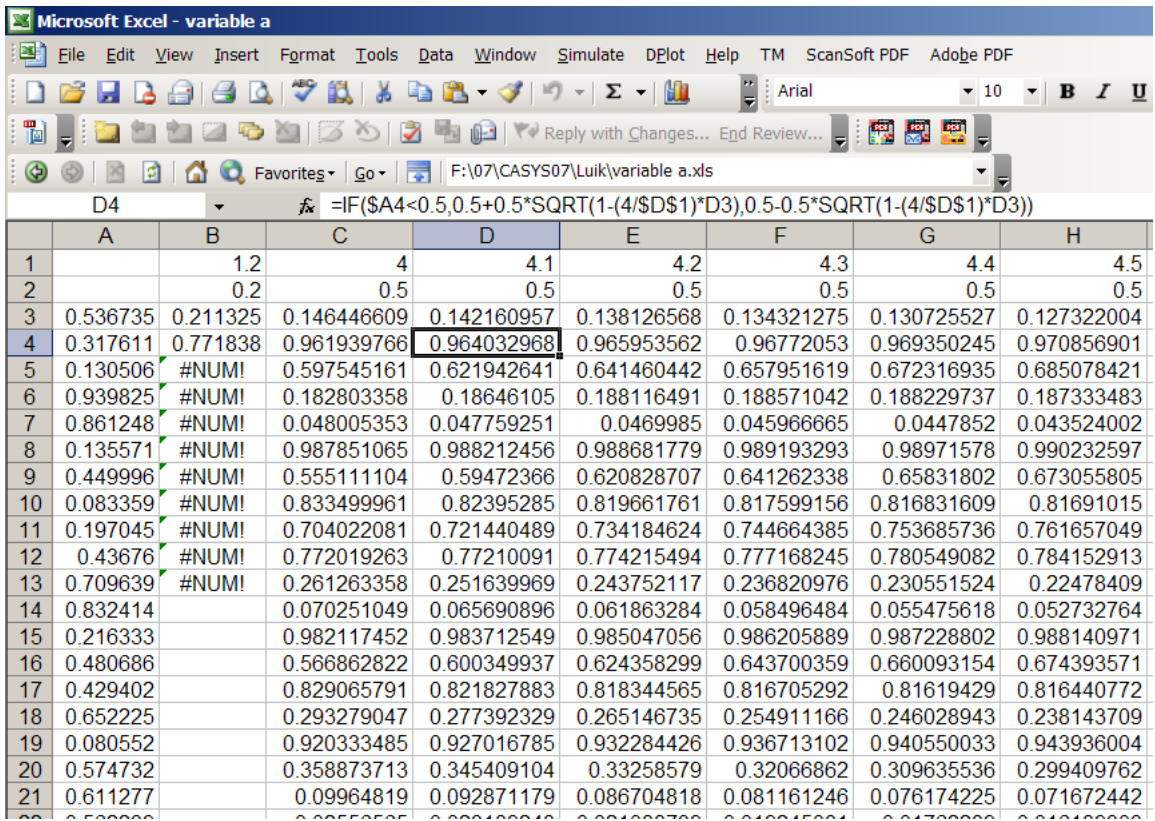

**FIGURE 4.** Simulation of the hyper-incursive equation (Equation 1) using MS Excel.

Figure 4 shows the simulation of Equation 1 for different values of *a* (in the first row). I have highlighted cell D4. The formula bar in the Excel sheet reads: if the random number in cell A4 (in this case, 0.317611) is smaller than 0.5, the positive sign of the square root is used and otherwise the negative one. The evaluation is based on the value of *a* in row one (D1 = 4.1), and the previous solution of the equation in the above cell (D3 = 0.142160957).

---

[4] Dubois [32: 208f.] defined "hyperincursion" as an incursion that generates multiple future states.

Note that for values of $a < 4$ (as in the B-column of Figure 3: $a = 1.2$), the number of steps is limited (as argued above).

In this simulation, the decisions were made randomly at each next time-step. This leads to simulation results as provided in Figure 5. Figure 5 shows that the system can sometimes dwell in a specific state. Next-order mechanisms like institutionalization of the rule for decision-making can be expected to stabilize these configurations because decisions are then no longer taken randomly [34].

Note that because of the structural coupling of the development of the social system as a strongly anticipatory system of meaning processing to decision-making by weakly anticipatory systems, the reduction of the uncertainty is provided by the incursive routines, and not by the hyper-incursive fluxes of communication as perhaps suggested by the original model of *autopoiesis* [21,35]. The communication of meaning is more complex than is assumed in the *autopoiesis* model which is biologically inspired and focuses on the self-organization of information processing. (The communication of meaning among the *models* will be addressed in a later section; at this level, it will be shown (below) that the self-organization of the interactions can also lead to a reduction of uncertainty.)

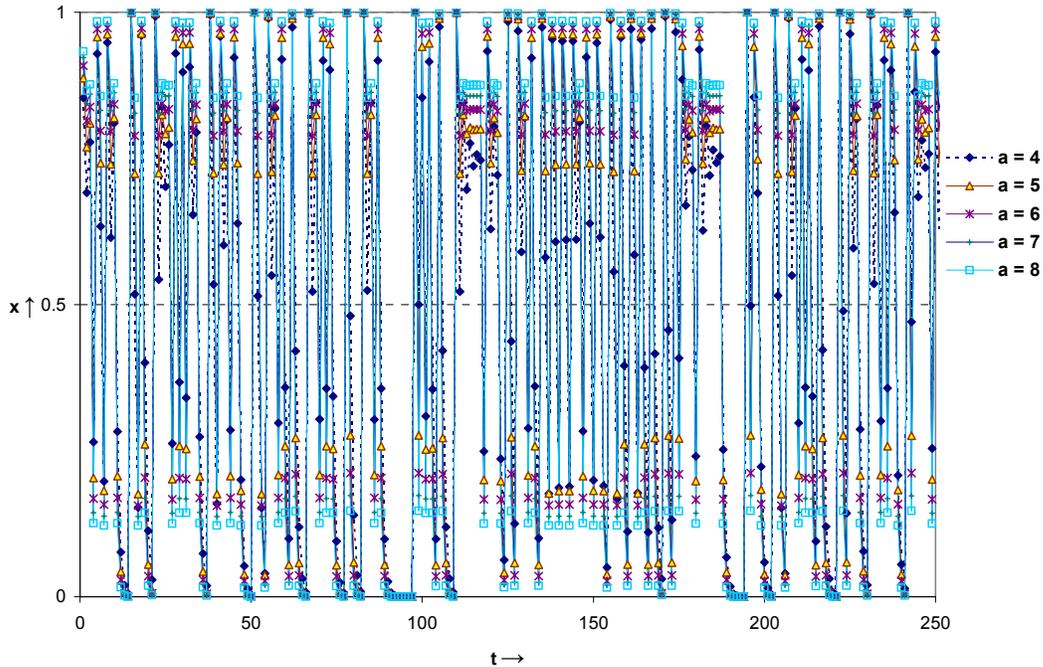

**FIGURE 5.** Trajectory of a social system based on random decisions of the decision-making units, for various values of the bifurcation parameter *a*.

The difference between structural coupling at the level of information processing in biological systems and the structural coupling between the incursive and hyper-incursive meaning processing in strongly anticipatory systems has been expressed in qualitative theorizing as the difference between structural coupling and interpenetration [3:437,5, 6:286 ff.,14:210 ff.]. Psychological and social systems are not only coupled structurally, but their operations in the second contingency—the intentional layer—are additionally coupled because each system can provide meaning to its own operation and reflexively to the meaning-processing in the structurally coupled one. When not only the operational systems but also their anticipatory models are structurally coupled, "interpenetration" between the systems makes it possible for the two systems to develop further by using each other's complex dynamics reflexively as relevant selection environments [28:174].

## 5. THE VARIOUS COUPLINGS OF MEANING-PROCESSING OPERATIONS

The logistic equation can be considered as a special case of the *perturbed recursion model* which Andersen [36:170 ff.] formulated as follows:

$$S_t = F(S_{t-1}, P) \qquad (6)$$

$F$ is a recursive function that transforms the state $S_{t-1}$ into a new state $S_t$, using a set of parameters $P = p_1, ..., p_n$.

Figure 6 shows the configuration when a system is continuously perturbed. The theory of *autopoiesis* further specifies that the system *S* itself makes a distinction between the parameters <$p_1...p_n$> (representing the environment) and its own state <$s_1...s_n$>. Since operationally closed, the system endogenously constructs and reproduces its boundaries. As I shall now show, these boundaries are constructed differently when a system has in addition to recursive routines also incursive ones available.

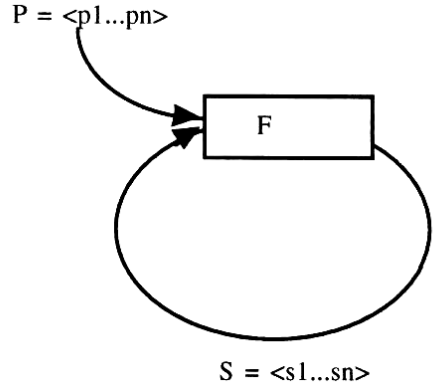

**FIGURE 6.** The perturbed recursion model. Source: [36:170].

Baecker [37:86 ff.] argued that in the case of operational closure and consequently *autopoiesis* the function (*F*) should be considered as an operator of the communication system (*S*) itself. Therefore, he reformulated Equation 6 as follows:

$$S = S(S,E) \tag{7}$$

This formulation leads to a paradox [38,39], but this paradox is a consequence of the static framework. Using the theory and computation of anticipatory systems, one can add time parameters to the different states and a subscript *c* (of *coding*; see below) to the operation as follows:

$$S_t = S_c(S_{t-1}, E) \tag{7a}$$

This rewrite of the functionality by the system itself means that the system must be able to develop and repair itself operationally in addition to reproducing itself [36:174]. The repair mechanism in communication is coding; the code of the communication enables a system to distinguish reflexively between signals and noise [40-42].

Following Parsons, Luhmann [6,14] added that codified meanings can be "condensed" in subsystems that are *functionally* differentiated. The functionality of the differentiation can be sustained if the subsystems are able to use the specificity of their respective codes for the reproduction of the system as a whole. Using the various codes, events can be appreciated differently within each subsystem; this diversity enables the system to process more complexity and can therefore be considered as functional to the further development of the system. The communications can then increasingly be refined. In the social system, for example, religious and scientific communications can be distinguished since the time of Galilei [43]. Similarly, political and scientific communications can be distinguished as based on providing different meanings to potentially the same events.

When functionality of the differentiation among different symbolically generalized codes of communications prevails—that is, in modern and pluriform societies—the external environment *E* is decomposed for each subsystem of communication into other subsystems, with a remaining term *ε* as representation of the residual environment. Let me use the lower-case *c* for the coding and *s* for this level of subsystems, and then rewrite Equation 7a as follows:

$$s_{i,t} = c_i(s_{i,t-1}, s_{j,t-1}, s_{k,t-1}, s_{l,t-1},..., \varepsilon) \tag{8}$$

In a functionally differentiated system, the windowing of the subsystems upon each other becomes horizontal. Each subsystem (*i*) codes ($c_i$) its own previous development and the development which it finds in its relevant environments. While the relation to the subsystem itself proceeds historically, that is, from its state at (*t* − 1) to its state at *t*, meaning is provided to the development in the other subsystems from the perspective of hindsight. Thus, we can obtain the additional equation:

$$s_{i,t} = c_i(s_{i,t-1}, s_{j,t}, s_{k,t}, s_{l,t}, ..., \varepsilon) \qquad (9)$$

The state of a subsystem ($s_{i,t}$)—and, therefore, of the system—is now dependent both on the previous state of this subsystem ($s_{i,t-1}$), and the current states of other subsystems. Equation 9 is incursive because the current state of another part of the same system is among the independent variables.

In other words, the functionally differentiated system contains an additional $\Delta t$ which can be used for a local reversal of the time axis and thus generate a transversal incursion which stands orthogonally to the longitudinal incursions provided by weak anticipation—using $x_{t+1} = ax_t (1 - x_{t+1})$. Weak anticipation along the longitudinal axis is available within all intentional (sub)systems, that is, both at the psychological and social levels [20,23]. For example, the development of one subsystem of society (e.g., a technology $i$) contains both a reference to a previous state of this technology and a reference to a current state of the selecting subsystem, e.g., the market $j$ [44].

The transversal incursion can be elaborated in a manner analogous to that of the incursive formulation of the logistic equation provided in Equation 5, but with an additional subscript for different subsystems:

$$x_{i,t} = ax_{i,t-1}(1 - x_{j,t}) \qquad (10)$$

On the basis of Equation 10 one can generalize to Equation 11 for cases where a variety of subsystems provide additional selection environments for the development of subsystem $i$, as follows:

$$x_{i,t} = ax_{i,t-1}(1 - x_{j,t})(1 - x_{k,t})(1 - x_{l,t})...\varepsilon \qquad (11)$$

Each subsystem ($i$) develops with reference to its own previous state, but one can expect all other subsystems to feedback upon this development as a relevant environment. Since each meaning-providing subsystem ($i$) also provides meaning to its own development longitudinally, and the two types of meaning-providing can interact, one can rewrite Equation 11 as follows:

$$x_{i,t} = ax_{i,t-1} \prod_{j=1; j \neq i}^{n}(1 - x_{j,t}) \cdot \varepsilon \qquad (12)$$

In Parsons's structural-functionalism, the number of subsystems ($n$) was analytically restricted—given his so-called four-function paradigm—but this number can vary in Luhmann's theory with the historical development of the media of communication and the symbolic generalization of specific codes. When the various subsystems use different frequencies for the updates, parameters have to be added to the corresponding selection mechanisms in Equation 12 and one obtains Equation 13:

$$x_{i,t} = ax_{i,t-1} \prod_{j=1; j \neq i}^{n} b_j (1 - x_{j,t}) \cdot \varepsilon \qquad (13)$$

The importance of the different frequencies of selections—and the accordingly different parameters $b$—for the further development of the system will become manifest below when we study the interactions between these selection terms in terms of both Equations 8 and 9, that is, recursively, incursively, and potentially hyper-incursively.

## 6. INTERACTION, ORGANIZATION, AND SELF-ORGANIZATION

Social order cannot be considered as a stable phenomenon because it contains an order of reproduced expectations in addition to being organized. When the expectations operate upon one another and interact, they can be expected to generate a non-linear dynamics that processes meaning. Specific meanings can be stabilized, for example, in social institutions, but all meaning arises from a horizon of possible meanings. As noted, Luhmann proposed that Husserl's original concept of intersubjectivity needed to be refined: in addition to interactions at the micro-level, communication of meaning can also be organized at the meso-level, and the fluxes of communication can be considered as self-organizing at the macro-level of society [45].

### 6.1. Interaction

While "double contingency" is a consequence of an internal model of relations between intentional systems, the interaction, organization, and self-organization of meaning can be considered as operations at the level of the models. These models operate selectively upon the systems which are modeled and upon each other. The interactions of the models therefore generate *selection environments* of possible meanings [20].

In Equation 13, I specified selections upon selections as the product of the selective factors. The hyper-incursive interaction between two incursive models can accordingly be modeled as follows:

$$x_t = b(1 - x_{t+1})(1 - x_{t+1}) \tag{14a}$$

$$x_t / b = 1 - 2x_{t+1} + x_{t+1}^2 \tag{14b}$$

$$x_{t+1}^2 - 2x_{t+1} + (1 - x_t / b) = 0 \tag{14c}$$

$$x_{t+1} = 1 \pm \sqrt{x_t / b} \tag{14d}$$

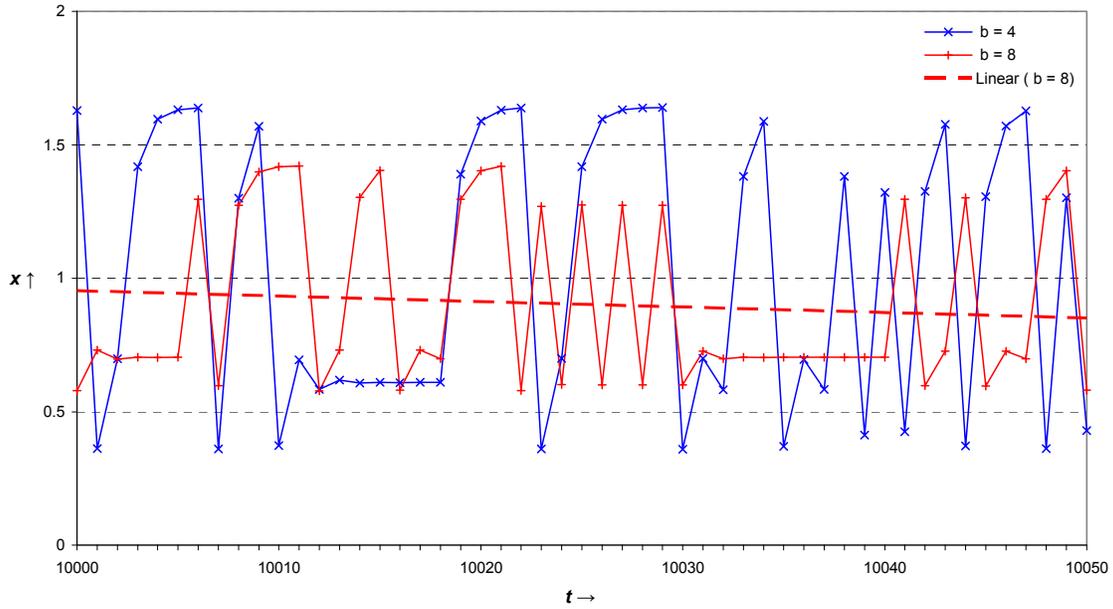

**FIGURE 7.** Hyper-incursive interaction.

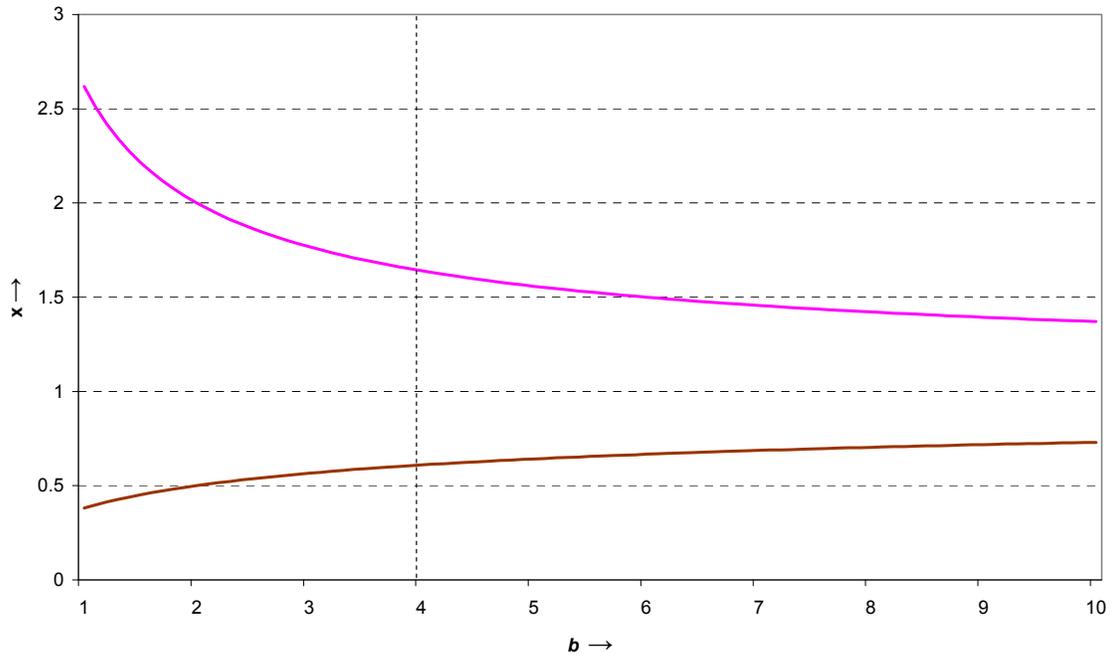

**FIGURE 8.** Two steady states of hyper-incursive interactions.

Figure 7 provides the simulation of this formula after 10,000 time-steps for $b = 4$ and $b = 8$, respectively. The auxiliary line is the linear regression line for $b = 8$ in this run. (The deviation from zero in the gradient is an effect of random fluctuations in choosing the plus or the minus signs in Equation 14d.) As expected, the interaction system swings from one side to the other, while sometimes staying on the same side for a number of iterations of the simulation. There is no progression in this system, nor can it come to an end.

Two steady states can be found by equating $x_{t+1}$ with $x_t$ in equation 14c:

$$x^2 - 2x + (1 - x/b) = 0 \tag{15a}$$

$$x^2 - (2 + 1/b)x + 1 = 0 \tag{15b}$$

$$x = (1 + \tfrac{1}{2b}) \pm \tfrac{1}{2b}\sqrt{(4b+1)} \tag{15c}$$

While approaching to a single steady for $b \to \infty$, the two steady states are continuous for variations in this parameter (Figure 8).

## 6.2. Organization

When a third term is added to Equation 14, we can obtain two formulas on the basis of Equations 8 and 9 above, respectively. I shall argue that this models the duality of structure which Giddens [26,46] specified in social theory [42]. Historical organizations of the interactions and potential self-organization among the fluxes of communication are modeled differently in Equations 16 and 17, respectively:

$$x_t = b(1 - x_{t+1})(1 - x_{t+1})(1 - x_t) \tag{16}$$

$$x_t = b(1 - x_{t+1})(1 - x_{t+1})(1 - x_{t+1}) \tag{17}$$

Equation 17 models what Strydom [47] called a "triple contingency": not only is there interaction, but a third party, i.e., a public, can be involved additionally. A public can only arise if a different code of communications was previously stabilized.

When three selections operate upon one another, a complex dynamics is generated [48,49]. For example, as long as the relationship is dyadic, Equations 14 and 16 prevail as models for the selection environment of meaning. However, when a child is added to a couple, or when the relationship is legalized as a marriage, another social system of expectations is generated (e.g., at the tax office). The development of this complex system is henceforth "non-trivial" [33], whereas *two* (sub)dynamics may co-evolve along a historical trajectory in a predictable process of mutual shaping. In addition to mutual relations, three subsystems can be expected to develop positions in their relations among one another [50] and thus to reduce the uncertainty in some configurations more than in others [19,51,52].

Equation 16 modifies Equation 17 by bending the self-organization of the fluxes back to the historically present state. The self-organizing system can thus provide itself with a retention mechanism for the results of interactions among intentional meaning-processing systems. Equation 16 can be developed analogously to Equation 14 into:

$$x_t = b(1 - x_{t+1})(1 - x_{t+1})(1 - x_t) \tag{16a}$$

$$x_{t+1}^2 - 2x_{t+1} + 1 - x_t/[b(1 - x_t)] = 0 \tag{16b}$$

$$x_{t+1} = 1 \pm \sqrt{x_t/b(1 - x_t)} \tag{16c}$$

Figure 9 shows that the organization of hyper-incursive interactions in historical time is always temporary. The light-blue curve (for $b = 10$), for example, shows that when the organization is not sufficiently changed by new rounds of interaction, the organization becomes unstable and fades out (in this case at $t = 6$). Technically, this originates from the factor $(1 - x_t)$ in the nominator under the square root of Eq. 16c: when $x_t \to 1$, this nominator becomes zero, and $x_{t+1}$ is no longer defined.[5] (The computer program MS Excel provides erroneously a zero value for these undefined states.) Unlike interactions and—as we shall see below—self-organizations of meaning in fluxes of communications, organizations remain historical. The other two mechanisms (interaction and self-organization) are needed to "revive" the historical layer of organization at various moments of time by providing variation and next-order stability, respectively.

---

[5] Daan van den Berg, *personal communication,* July 12, 2008.

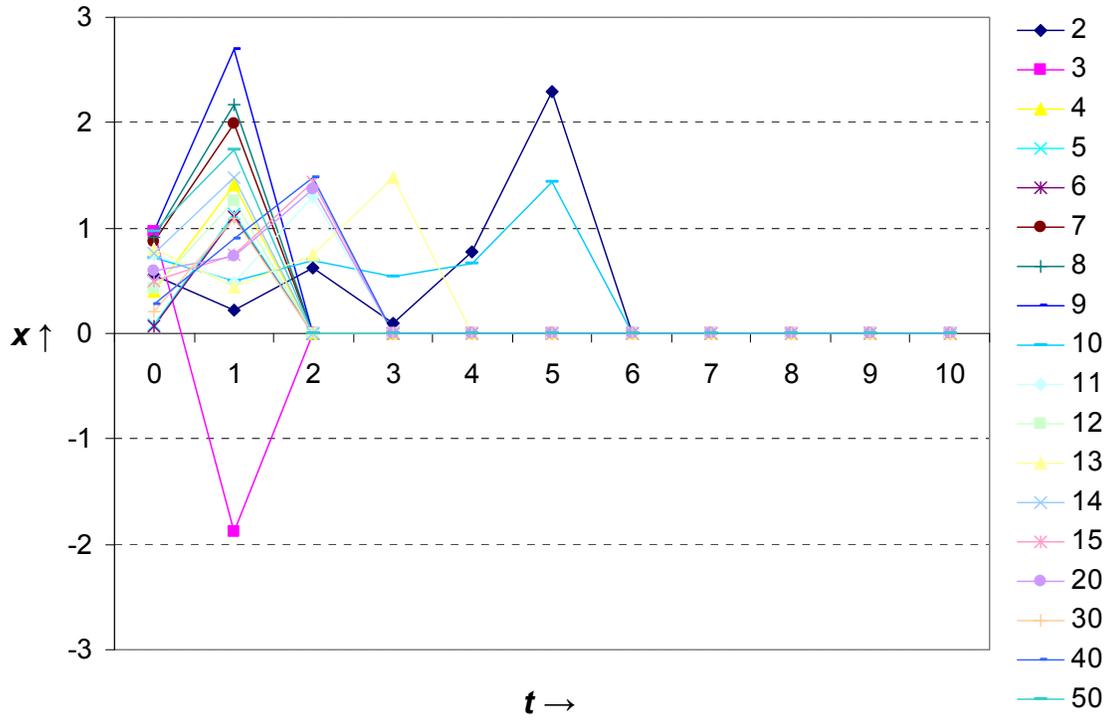

**FIGURE 9.** Organization of interactions for different values of the parameter *b*.

### 6.3. Self-organization

Equation 17 can be developed as follows:

$$x_t = b(1-x_{t+1})(1-x_{t+1})(1-x_{t+1}) \tag{17a}$$

$$\frac{x_t}{b} = (1-x_{t+1})^3 \tag{17b}$$

$$\left(\frac{x_t}{b}\right)^{1/3} = 1 - x_{t+1} \tag{17c}$$

$$x_{t+1} = 1 - \left(\frac{x_t}{b}\right)^{1/3} \tag{17d}$$

This equation has three roots of which two are complex. The real solution of 9d can (conventionally) be denoted as:

$$x_{t+1} = 1 - \sqrt[3]{\frac{x_t}{b}} \tag{18a}$$

and the complex solutions are:

$$x_{t+1} = 1 - \sqrt[3]{\frac{x_t}{b}\left(\frac{-1 \pm i\sqrt{3}}{2}\right)}$$
(18b)

Since the further operation cannot evaluate the complex solutions in a next time-step, the system in this case has to continue with the real solution. This leads to a single and stable solution for each value of *b*. Are these perhaps Luhmann's [53:e.g., 394, 1127] "eigenvalues"?

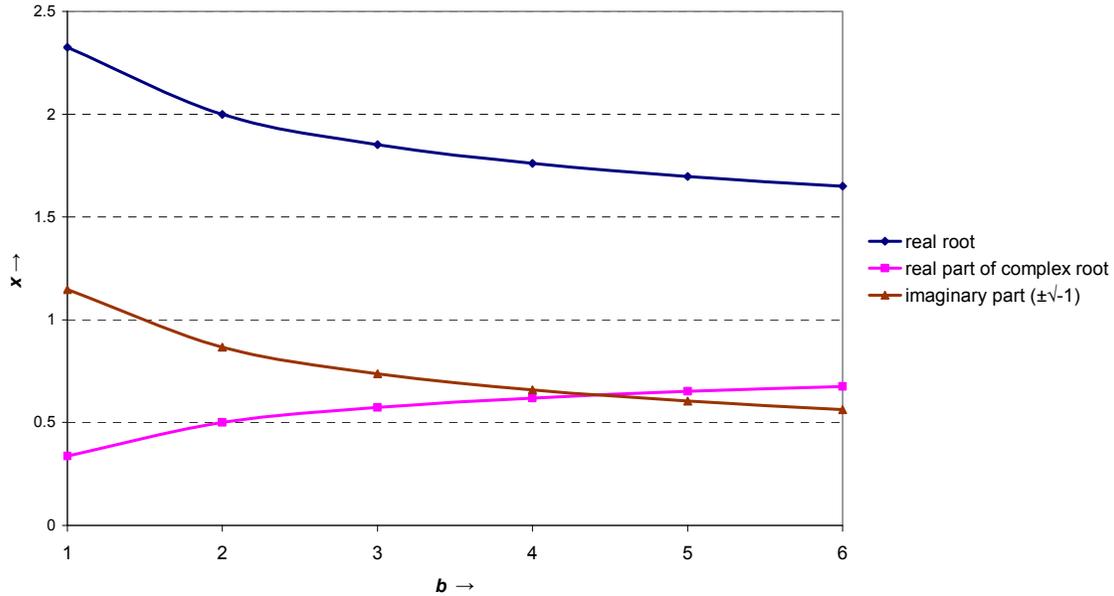

**FIGURE 10.** Real and complex roots of the cubic equation in the case of three selections operating upon one another.

Figure 10 provides the steady state values for different values of the parameter *b*. The curve of the imaginary part goes asymptotically to zero. Changes in *b*—because of the concurrent operation of organizing meaning in historical time by using Equation 16—can move the system along the lines in Figure 8. At each moment, however, the system quickly converges on a single steady state.

My thesis [19] is that in a modern and pluriform society which is functionally differentiated in terms of (symbolically generalized) codes of communication in subsystems, the recombination and temporary reintegration of these codes in organizations can change the system by reorganizing it (e.g., in a techno-economic co-evolution). Therewith the value of the parameter *b* is potentially changed in a next round, and the self-organization in the horizon of meanings can be expected to change accordingly. Thus, the two layers of the social system can auto-catalytically drive each other while controlling the uncertainty.

## 7. DISCUSSION

The simulation of organizations showed that the organization of meaning exchange is temporary. Interactions are continuously providing variation; self-organization structures the system of meaning exchanges in terms of codes of communication. The self-organization of meaning in communication provided a single solution for all values of the parameter (*b*). However, the system and subsystems are both concurrent and nested: under the condition of functional differentiation among the subsystems, each subsystem can be expected to provide its own solution (eigenvalue *b'*) for the self-organization when observed from the system's level. Figure 11 [23:411] provides a visualization for this nesting of intra- and intersystem anticipation.

Change comes from the different parameter values at the *sub*systems level which respectively converge on different roots of the cubic equations. These values are recombined historically by a specific *organization* at the meso-level of the (next-order) system. Using Luhmann's theory, one can associate the specificity of the various constants (that is, the converging roots of the cubic equations) with the functionally differentiated codes of communication in the respective subsystems.

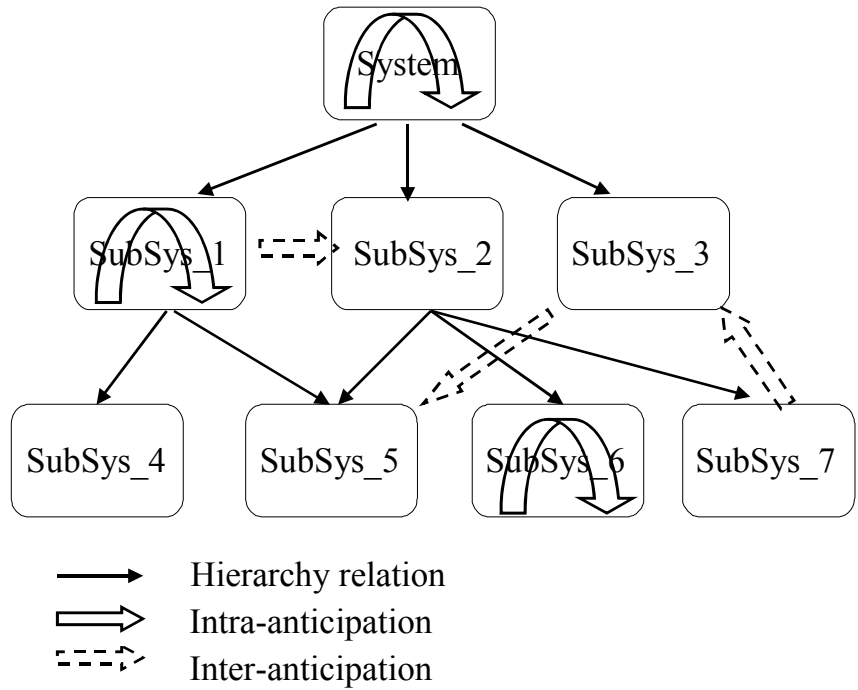

**FIGURE 11.** Intra and Inter-Anticipation. Source: [23:411].

The differentiation among the codes is not only a static differentiation among eigenvectors, but also a differentiation in terms of frequencies [43]. For example, a political system can be expected to operate with an election cycle of four years, while science operates in terms of quarterly journals and annual meetings. The economy updates with prices and stock markets on an almost daily basis.

Luhmann's social systems theory was hitherto mainly formalized using Spencer Brown's (1969) *Laws of Form* [54,55]. Spencer Brown distinguished between two states: the 'marked' and the 'unmarked.' Luhmann [e.g., 38]— and many others—have believed that this distinction could be used for the development of a general calculus [39,56,57]. However, Spencer Brown's marks are made at specific moments of time. Their recursive elaboration leads to a topology [39,55], but not to a calculus.

In a later discussion, Spencer Brown [58:51] added the possibility of an evolutionary process, but he formulated this mathematically as only an oscillation between the two states. An oscillation can, for example, be produced by the logistic equation ($3 \leq a < 4$). However, my algorithmic equations were specified on the basis of substantive theorizing about the communication of meaning in social system [cf. 56:20,59]. In other words, the evolutionary theorist is not interested only in the momentary existence of an observer or a second-order observer, but also in developments over time in terms of specific operations [57,59,60].

## 8. CONCLUSIONS

In this study, I constructed a relation between Dubois' incursive and hyper-incursive equations of strongly anticipatory systems [1,8,9] and Luhmann's social systems theory [6,14]. I used both theories heuristically and perhaps eclectically because the construction of a relation requires a creative translation of the one theory into the other. Luhmann [6:605, 14:446f.], for example, formulated that "(s)elf-referential, autopoietic reproduction would not be possible without an anticipatory recursivity." However, Luhmann did not specify this mechanism.

In his summarizing *chef-d'oeuvre* of 1997 [53], Luhmann provided two references to Robert Rosen's *Anticipatory Systems* [17], but both were framed in a biological context [53: 206 and 820]. He added (at p. 821) that in the domain of meaningful processing of information one cannot avoid a reference to the present when defining the relation between past and future; consequently, time has to be conceptualized as a dimension. Luhmann's time-dimension with the present at its origin can further be elaborated as a degree of freedom in the theory and computation of anticipatory systems.

In other contexts, Luhmann developed a semantics for the discussion of "time" [61: 98, n.10], but for my argument other elements of social-systems theory are more important than these historical elaborations. Luhmann [53] draws two main axes of differentiation within social systems: social differentiation and systems differentiation. Social differentiation in the communication is possible because communication can be coded differently and codes of communication can be generalized symbolically. For example, one can code communication economically, scientifically or politically. Money, power, validity, affection, trust, etc., can be used to shortcut communication and thus to process more complexity [62,63]. The different function systems operate in parallel, but not synchronously. As Luhmann formulated: "Social differentiation serves as an uncoupling mechanism. It divides the time-orientations in the different systems and therefore accepts that things can be urgent in one system, while another system can take its time."[61:128]

Systems differentiation organizes the social system at different levels. Face-to-face communication or interaction among incursive systems remains the micro-operation for strong anticipation in the processing of meaning. Hyper-incursion models the double contingency which is naturally given among reflexive agents. However, communications can also be organized at the meso-level.[6] At the macro-level of society, Luhmann [12] proposed to use Maturana & Varela's [35] concept of *autopoiesis* or self-organization: interhuman coordination is possible because meaning can also be communicated (e.g., using languages or symbolic codes of communication). The differentiation of micro-level interaction, meso-level organization, and macro-level self-organization in the communication of meaning was elaborated in terms of different equations which represent the selection environments of intentional systems. Some meanings can be retained in the knowledge base of this system as more meaningful than others.

The model of recursive perturbations allowed us to consider both Spencer Brown's formulae and the various formulations of the logistic map as species of a family of equations. The spanning of the time dimension enables us to dissolve the paradoxes in social-systems theory (as in the above case of Equation 7). The theory and computation of anticipatory systems further develops the time dimension into a degree of freedom which can be studied in the forward and/or backward mode. The duality of structure in the social system [26,46] could thus be considered as the result of a difference in the time-subscripts of Equations 16 and 17 (above). Note that these equations are no longer exclusively based on the model of recursive perturbations or the logistic map. In addition to the bifurcation parameter $a$, a second parameter $b$ is involved in modeling the non-linear dynamics of meaning-processing at the supra-individual level.

In summary: Leydesdorff & Dubois [22] showed that the social system can be considered as inherently anticipatory because of its distributedness at each moment in time. This distribution can be organized in terms of a functional differentiation among subsystems which can operate asynchronously and thus generate another $\Delta t$. Reflexivity and intentionality add a second (longitudinal) anticipatory mechanism to all intentional (sub)systems. The two mechanisms can be combined into a strongly anticipatory mechanism of meaning-processing [20]. A strongly anticipatory system co-constructs its next stages as one of its subdynamics. The social system provides us with a natural example of such a system.

Marx already in 1867 expressed this expectation of a dynamics at the supra-individual level as the autonomous accumulation of capital [64]. Nowadays, we are able to refine this accumulation as the fragile progression of a knowledge-based economy [19].

## ACKNOWLEDGMENTS

I am grateful to Dirk Baecker, Mark Burke, Sander Franse, Daniel M. Dubois, Stig Holmberg, Juan Jesús Torres Carbonell, and Daan Van den Berg for suggestions and communications in previous stages of this paper.

---

[6] Organization and decision-making can also be considered as functions of the social system. Decision-making, for example, can be codified into decision rules. I elaborate this in [19: 147 ff].